\documentclass[showpacs,preprintnumbers,amsmath,amssymb,twocolumn]{revtex4}

\usepackage{graphicx}
\usepackage{dcolumn}
\usepackage{bm}

\begin{document}

\title{\bf\large{Chiral Phase Transition beyond Mean Field Approximation}}
\author{Lianyi He and Pengfei Zhuang}
\affiliation{Physics Department,
Tsinghua University, Beijing 100084, China}

\begin{abstract}
Based on the analogy between the Nambu--Jona-Lasinio model of
chiral symmetry breaking and the BCS theory of superconductivity,
we investigate the effect of $\bar q q$ pair fluctuations on the
chiral phase transition. We include uncondensed $\bar q q$ pairs
at finite temperature and chemical potential in a self-consistent
T-matrix formalism, the so-called $G_0 G$ scheme. The pair
fluctuations reduce significantly the critical temperature and
make quarks massive above the critical temperature.
\end{abstract}

\date{\today}
\maketitle

It is well known that the spontaneous breaking of chiral symmetry
is one of the most important features of the vacuum of Quantum
Chromodynamics (QCD). The spontaneous breaking successfully
explains many low energy phenomena in QCD, such as the large
continues quark mass and the small pion mass. The quark-antiquark
condensate $\langle\bar q q\rangle$ which characters the
spontaneous breaking is about $(250\ $MeV$)^3$ from the QCD sum
rule and lattice calculations. It is generally believed that, in
hot and dense medium, the chiral condensate will be reduced and
the broken chiral symmetry will be restored at sufficiently high
temperature and density.

Since the chiral symmetry breaking and restoration happen in the
non-perturbative region, it is hard to handle them directly by the
original QCD lagrangian. The investigation relies mainly on low
energy effective models and lattice calculations. One of the
successful models is the Nambu--Jona-Lasinio (NJL)
model~\cite{NJL1,NJL2}. The idea of the model is from the
well-known Bardeen-Cooper-Shriffer (BCS) theory of
superconductivity. In superconducting metals, the Cooper pairing
between electrons in the spin-singlet state leads to a condensate
$\langle\psi_\uparrow\psi_\downarrow\rangle$ which spontaneously
breaks the electromagnetic U$(1)$ symmetry and causes an energy
gap. In analogy, the pairing between a quark and an antiquark with
the same chirality leads to a condensate $\langle\bar q q\rangle$.
Such a condensate spontaneously breaks the chiral symmetry and
gives a dynamical mass to the quarks.

While the analogy between the QCD and BCS vacua leads to a
successful theory of chiral symmetry breaking, one should keep in
mind that there is an important difference between the two: In
contrast to the BCS superconductivity, the chiral symmetry
breaking is a strong coupling phenomenon. From the recent studies
on relativistic heavy ion collisions, there may exist a strongly
coupled quark-gluon plasma phase (sQGP)~\cite{SQGP} above the
critical temperature $T_c$ of the deconfinement and chiral phase
transitions where both quarks and their bound states are
constituents of the system. From lattice simulations, any
thermodynamic quantity can not reach its Stefan-Boltzmann limit
even at extremely high temperature~\cite{SB}. A possible
explanation for this thermodynamic suppression is that quarks have
a large thermal mass in the deconfined and chiral restored
phase~\cite{SQGP,LAT,HTL}.

In the language of condensed matter physics, these phenomena
strongly indicate that there exists a pseudogap for
quarks~\cite{qqbar,qqbar2,zarebo, kunihiro} and the matter may be
in the crossover region from BCS to Bose-Einstein condensation
(BEC)~\cite{BCSBEC1,BCSBEC2,BCSBEC3}. It is generally believed
that the BCS mean field theory describes well the crossover at
zero temperature. However, to obtain a quantitatively correct
$T_c$ and a correct fermion excitation above $T_c$, one should
properly consider pair fluctuations at finite temperature. There
are some T-matrix schemes to treat the finite temperature pair
fluctuations. Among them, the self-consistent pair approximation
or the so-called $G_0 G$ scheme~\cite{BCSBEC3,G0G1,G0G2} is a
competitive one. The effect of the pair fluctuations in this
scheme is treated as a pesudogap. Recently, the $G_0 G$ theory is
applied to the study of color superconductivity in the NJL
model~\cite{he}. In this paper, we will investigate the chiral
phase transition in the NJL model, by using the self-consistent
$G_0 G$ theory.

To have a BCS-like description of the spontaneous chiral symmetry
breaking, it is useful to take the two-dimension Wely spinors
defined as
\begin{equation}
q=\left(\begin{array}{cc} q_{\text L}\\
q_{\text R}\end{array}\right),\ \ \bar{q}=\left(\begin{array}{cc}
q_{\text R}^\dagger& q_{\text L}^\dagger\end{array}\right)
\end{equation}
where the color and flavor indexes of the quark field are not
explicitly shown. The kinematic term of the NJL Lagrangian, ${\cal
L}_0=\bar{q}i\partial\!\!\!\!/q$, can be written as
\begin{equation}
{\cal L}_0=q_{\text L}^\dagger
i(\partial_t-\mbox{\boldmath{$\sigma$}}\cdot\mbox{\boldmath{$\nabla$}})q_{\text
L}+q_{\text R}^\dagger
i(\partial_t+\mbox{\boldmath{$\sigma$}}\cdot\mbox{\boldmath{$\nabla$}})q_{\text
R},
\end{equation}
where $\sigma_i$ are the Pauli matrices. The interaction between
quarks and antiquarks in the NJL model are expressed by a four
fermion coupling term, it can be generally written as
\begin{equation}
{\cal L}_{\text I}
=\frac{g}{4}\left[\left(\bar{q}q\right)^2+\left(\bar{q}i\gamma_5\Gamma
q\right)^2 \right],
\end{equation}
where $\Gamma=1$ and $\mbox{\boldmath{$\sigma$}}$ are,
respectively, for one and and two flavor cases. To see directly
the analogy between the NJL model of chiral symmetry breaking and
the BCS theory of superconductivity, it is intuitive to look into
the one flavor case. Using the identities for the Wely spinors,
\begin{eqnarray}
\bar{q}q&=&q_{\text R}^\dagger q_{\text L}+q_{\text L}^\dagger q_{\text R}\nonumber\\
\bar{q}i\gamma_5q&=&-i(q_{\text R}^\dagger q_{\text L}-q_{\text
L}^\dagger q_{\text R}),
\end{eqnarray}
the four fermion interaction can be written as
\begin{equation}
{\cal L}_{\text I} = gq_{\text R}^\dagger q_{\text L}q_{\text
L}^\dagger q_{\text R},
\end{equation}
which has the same structure of the interaction between fermions
in BCS theory where one has ${\cal L}_{\text
I}=g\psi_\uparrow^*\psi_\downarrow^*\psi_\downarrow\psi_\uparrow$.

The order parameter field describing chiral symmetry breaking can
be defined as $\Phi(t,{\bf x})=-gq_{\text R}^\dagger q_{\text L}$.
The QCD vacuum is characterized by the chiral condensate $\langle
\Phi\rangle$ which means the pairing between a quark and an
antiquark with the same chirality. Since the vacuum has a definite
parity, the chiral condensate is a real number, $\langle q_{\text
R}^\dagger q_{\text L}\rangle=\langle q_{\text L}^\dagger q_{\text
R}\rangle$.

We start by rewriting the mean field theory of the two flavor NJL
model in a T-matrix formalism. Such a formalism is important for
us to go beyond the mean field by including the non-condensed
$\bar q q$ pairs at finite temperature. The mean field quark
propagator ${\cal S}$ reads
\begin{equation}
{\cal S}^{-1}(k)= \left(\begin{array}{cc} {\cal G}_{0\text{L}}^{-1}(k)&-m_{\text{sc}}\\
-m_{\text{sc}}&{\cal G}_{0\text{R}}^{-1}(k)\end{array}\right),
\end{equation}
where ${\cal
G}_{0\text{L},\text{R}}^{-1}(k)=i\nu_n+\mu\pm\mbox{\boldmath{$\sigma$}}\cdot{\bf
k}$ are the free quark propagators with $\nu_n$ being the fermion
frequency $\nu_n=(2n+1)\pi T\ (n=0,\pm 1,\pm 2,\cdots)$, and
$m_{\text{sc}}=\langle \Phi\rangle$ is the order parameter of the
phase transition. The quark chemical potential $\mu$ is introduced
by considering conserved charge density $q^\dagger q$. The
explicit form of the quark propagator can be expressed as
\begin{equation}
{\cal S}(k)= \left(\begin{array}{cc} {\cal G}_{\text{L}}(k)&{\cal F}_{\text{L}}(k)\\
{\cal F}_{\text R}(k)&{\cal G}_{\text{R}}(k)\end{array}\right)
\end{equation}
with the matrix elements given by
\begin{eqnarray}
\label{propagator}
{\cal G}_{\text{L},\text{R}}(k)&=&\left[{\cal
G}_{0\text{L},\text{R}}^{-1}(k)-\Sigma_{\text{L},\text{R}}(k)\right]^{-1},\nonumber\\
{\cal F}_{\text{L},\text{R}}(k)&=&m_{\text{sc}}{\cal
G}_{\text{L},\text{R}}(k){\cal G}_{0\text{R},\text{L}}(k),
\end{eqnarray}
where the self-energies $\Sigma_{\text{L,R}}$ at mean field level
are defined as
\begin{equation}
\label{selfenergy}
\Sigma_{\text{L},\text{R}}(k)=m_{\text{sc}}^2{\cal
G}_{0\text{R},\text{L}}(k).
\end{equation}
After some simple algebra, the diagonal and off-diagonal elements
can be evaluated as
\begin{eqnarray}
{\cal
G}_{\text{L},\text{R}}(k)&=&\frac{i\nu_n+\mu\mp\mbox{\boldmath{$\sigma$}}\cdot{\bf
k}}{(i\nu_n+\mu)^2-{\bf k}^2-m_{\text{sc}}^2},\nonumber\\
{\cal
F}_{\text{L},\text{R}}(k)&=&\frac{m_{\text{sc}}}{(i\nu_n+\mu)^2-{\bf
k}^2-m_{\text{sc}}^2}.
\end{eqnarray}
For those who are familiar with the BCS theory of
superconductivity, ${\cal G}$ and ${\cal F}$ are in analogy to the
normal and anomalous Green functions, and the quark chemical
potential plays the role of effective Zeeman splitting between
quarks and antiquarks. At $\mu=0$, the quark propagator has two
poles $i\nu_n=\pm\sqrt{{\bf k}^2+m_{\text{sc}}^2}$, which means
that quarks obtain a mass gap $m_{\text{sc}}$ in the chiral
symmetry breaking phase. The order parameter $m_{\text{sc}}$ is
determined by the self-consistent gap equation
\begin{equation}
\label{gap1} m_{\text{sc}}=-g\sum_{k}\text{Tr}{\cal
F}_{\text{L},\text{R}}(k),
\end{equation}
where the summation over quark momentum is defined as
$\sum_{k}=T\sum_n\int d^3{\bf k}/(2\pi)^3$ and the trace is taken
in color and flavor spaces.

In the mean field theory, $\bar q q$ pairs enter into the problem
only through the condensate. The condensed pairs can be associated
with a T-matrix in such a way
\begin{equation}
t_{\text{sc}}(q)=\frac{m_{\text{sc}}^2}{T}\delta(q)
\end{equation}
that the quark self energies (\ref{selfenergy}) can be formally
expressed as
\begin{equation}
\Sigma_{\text{L},\text{R}}(k)=\sum_{q}t_{\text{sc}}(q){\cal
G}_{0\text{R},\text{L}}(k-q).
\end{equation}

From the gap equation (\ref{gap1}), the mean field theory is
related to a particular asymmetric pair
susceptibility~\cite{BCSBEC3},
\begin{eqnarray}
\label{susceptibility}
\chi(q)&=&\chi_{\text{LR}}(q)=\chi_{\text{RL}}(q)\nonumber\\
&=&\frac{1}{2}\sum_{k}\text{Tr}\Big[{\cal
G}_{\text{L}}(k){\cal G}_{0\text{R}}(k-q)\nonumber\\
&&+{\cal G}_{\text{L}}(k-q){\cal G}_{0\text{R}}(k)\Big],
\end{eqnarray}
from which the gap equation in the symmetry breaking phase is
given by the condition
\begin{equation}
\label{gap0}
1+g\chi(0)=0.
\end{equation}
This suggests that the uncondensed $\bar q q$ pair propagator
should be in the form
\begin{equation}
t_{\text{pg}}(q)=\frac{g}{1+g\chi(q)}.
\end{equation}
The gap equation (\ref{gap0}) is just the so-called BEC condition
$t_{\text{pg}}^{-1}(q=0)=0$. While the uncondensed $\bar{q} q$
pairs play no role in the BCS mean field theory, such a specific
choice of the pair susceptibility and the BEC condition are the
fundamental criterion for us to go beyond the mean field.

We now take into account the uncondensed $\bar q q$ pairs. In the
mean field theory, the quark self-energies
$\Sigma_{\text{L},\text{R}}$ contain only the contribution from
the condensed $\bar{q}q$ pairs. This is mostly correct at zero
temperature. However, at finite temperature, the condensed pairs
with zero total momentum can be thermally excited. Therefore, the
total contribution to the quark self-energies should include both
the condensed (sc) pairs and the uncondensed or
``pseudogap"-associated (pg) pairs\cite{BCSBEC3},
\begin{equation}
\Sigma_{\text{L},\text{R}}(k)=\sum_{q} t(q){\cal
G}_{0\text{R},\text{L}}(k-q)
\end{equation}
with the total pair propagator $t(q)$ defined by
\begin{eqnarray}
t(q)&=&t_{\text{pg}}(q)+t_{\text{sc}}(q),\nonumber\\
t_{\text{pg}}(q)&=&\frac{g}{1+g\chi(q)},\ \ \ q\neq0,\nonumber\\
t_{\text{sc}}(q)&=&\frac{m_{\text{sc}}^2}{T}\delta(q).
\end{eqnarray}
With the modified self-energies, the dressed quark propagators
${\cal G}_{\text{L},\text{R}}(k)$ and the pair susceptibility
$\chi(q)$ are still given by (\ref{propagator}) and
(\ref{susceptibility}). The beyond-BCS effect is reflected in the
quark self-energies $\Sigma_{\text{L,R}}$. The BEC condition
$t_{\text{pg}}^{-1}(0)=0$ and the quark self-energies form in
principle a coupled set of equations to determine the new order
parameter $m_{\text{sc}}$.

The above equations are hard to handle analytically. In the
symmetry breaking phase with $T\leq T_c$, the BEC condition
$t_{\text{pg}}^{-1}(0)=0$ implies that $t_{\text{pg}}(q)$ is
strongly peaked at $q=0$. This allows us to
approximate\cite{BCSBEC3}
\begin{equation}
\label{app}
\Sigma_{\text{L},\text{R}}(k)\simeq m^2{\cal
G}_{0\text{R},\text{L}}(k),
\end{equation}
where $m^2$ includes the contribution from both the condensed and
the thermally excited $\bar{q}q$ pairs,
\begin{equation}
m^2=m_{\text{sc}}^2+m_{\text{pg}}^2,
\end{equation}
with the pseudogap $m_{\text{pg}}$ defined as
\begin{equation}
\label{pg} m_{\text{pg}}^2 = \sum_{q\neq 0}t_{\text{pg}}(q).
\end{equation}
Note that, above the critical temperature $T_c$, the BEC condition
disappears and then such an approximation is in principle not
valid. How to practically treat the pair fluctuations and their
effect on the quark propagator above $T_c$ is still an open
question in the $G_0 G$ scheme. However, for temperatures not much
higher than $T_c$, this approximation may be still good to give
qualitatively correct result.

Under the approximation, the dressed quark propagator can be
analytically evaluated as
\begin{equation}
{\cal
G}_{\text{L},\text{R}}(k)=\frac{i\nu_n+\mu\mp\mbox{\boldmath{$\sigma$}}\cdot{\bf
k}}{(i\nu_n+\mu)^2-{\bf k}^2-m^2},
\end{equation}
which means that, when one goes beyond the mean field, quarks
obtain a mass gap $m=(m_{\text{sc}}^2+m_{\text{pg}}^2)^{1/2}$
rather than $m_{\text{sc}}$. One may worry about that the
appearance of the pseudogap $m_{\text{pg}}$ explicitly breaks the
chiral symmetry. Parallel to the discussion in the BCS-BEC
crossover\cite{G0G2}, we can show that $m_{\text{pg}}^2$ is just
the classical fluctuation of the order parameter field $\Phi$,
\begin{equation}
m_{\text{pg}}^2=\langle|\Phi|^2\rangle-\langle|\Phi|\rangle^2,
\end{equation}
and hence does not break the chiral symmetry.

After introducing the pesudogap $m_{\text{pg}}$, the pair
susceptibility can be analytically evaluated as
\begin{eqnarray}
\chi(q)&=&N_cN_f\sum_{e=\pm}\int\frac{d^3{\bf
k}}{(2\pi)^3}\frac{|{\bf k}-{\bf q}|+eE_{\bf
k}}{(i\omega_n)^2-(|{\bf
k}-{\bf q}|+eE_{\bf k})^2}\nonumber\\
&&\left[1+e\frac{|{\bf k}|}{E_{\bf k}}\cos\phi_{\bf
q}\right]\left[1-\bar{f}(eE_{\bf k})-\bar{f}(|{\bf k}-{\bf
q}|)\right],\nonumber\\
\end{eqnarray}
where $N_c$ and $N_f$ are color and flavor numbers, and the pair
frequency $\omega_n$, quark energy $E_{\bf k}$, the angle
$\phi_{\bf q}$ and the function $\bar f(x)$ are defined as
$\omega_n=2n\pi T\ (n=0,\pm 1, \pm 2,\cdots), E_{\bf k}=\sqrt{{\bf
k}^2+m^2}, \cos\phi_{\bf q}={\bf k}\cdot({\bf k}-{\bf q})/(|{\bf
k}||{\bf k}-{\bf q}|)$ and $\bar
f(x)=\left[f(x-\mu)+f(x+\mu)\right]/2$ with
$f(x)=1/\left(e^{x/T}+1\right)$ being the Fermi-Dirac distribution
function. The BEC condition $1+g\chi(0)=0$, namely the gap
equation can now be explicitly written as
\begin{equation}
\label{gap2} 1-gN_cN_f\int\frac{d^3{\bf
k}}{(2\pi)^3}\frac{1-2\bar{f}(E_{\bf k})}{E_{\bf k}}=0.
\end{equation}
The equations (\ref{pg}) and (\ref{gap2}) form a closed set to
determine the order parameter $m_{\text{sc}}$ and the pseudogap
$m_{\text{pg}}$ as functions of temperature and chemical
potential.

Solving such a coupled set of equations is still rather
complicated. In the temperature region below and around the
critical temperature $T_c$ which is expected to be much smaller
than the momentum cutoff $\Lambda$ of the NJL model, we can employ
the pole approximation for the pair propagator $t_{\text{pg}}$,
\begin{equation}
t_{\text{pg}}(\omega,{\bf
q})\simeq-\frac{Z^{-1}}{\omega^2-\omega_{\bf q}^2}
\end{equation}
with $\omega_{\bf q}^2=v^2{\bf q}^2+\Delta^2$, where we have taken
the analytical continuation $i\omega_n\rightarrow\omega+i0^+$, and
the coefficients $Z$,$v^2$ and $\Delta^2$ are defined as
\begin{eqnarray}
&&Z=-\frac{1}{2}\frac{\partial^2\chi}{\partial
\omega^2}\Big|_{\omega={\bf q}=0},\ \ \
v^2=\frac{1}{2Z}\frac{\partial^2\chi}{\partial {\bf
q}^2}\Big|_{\omega={\bf q}=0}\nonumber\\
&& \Delta^2=\frac{1}{gZ}\left[1+g\chi(0)\right].
\end{eqnarray}
The pair width is safely neglected below and around
$T_c$\cite{BCSBEC3}. With the help of the pole approximation, the
pseudogap equation (\ref{pg}) takes the simple form
\begin{equation}
\label{mpg}
m_{\text{pg}}^2=\frac{1}{Z}\int\frac{d^3{\bf
q}}{(2\pi)^3}\frac{1+2b(\omega_{\bf q})}{2\omega_{\bf q}},
\end{equation}
where $b(x)=1/\left(e^{x/T}-1\right)$ is the Bose distribution
function. Since the vacuum is well described by the mean field
theory, we require $m_{\text{pg}}=0$ at zero temperature and
chemical potential. To this end, we simply eliminate the first
term on the righthand side of (\ref{mpg}).

Before numerical calculations, we can analytically arrive at the
following conclusions:
\\ 1)In the symmetry breaking phase, the gap equation $1+g\chi(0)=0$ leads to $\Delta^2=0$,
and then the pair dispersion is gapless in the long wavelength
limit. At sufficiently low temperature, the pseudogap
$m_{\text{pg}}$ is proportional to the temperature $T$.
\\ 2)From $\chi, Z \sim N_c$, we have $m_{\text{pg}}^2 \sim 1/N_c$. In the large $N_c$ limit, the
pseudogap effect can be neglected, and the mean field
approximation describes well the chiral phase transition.

\begin{figure}[!htb]
\begin{center}
\includegraphics[width=7cm]{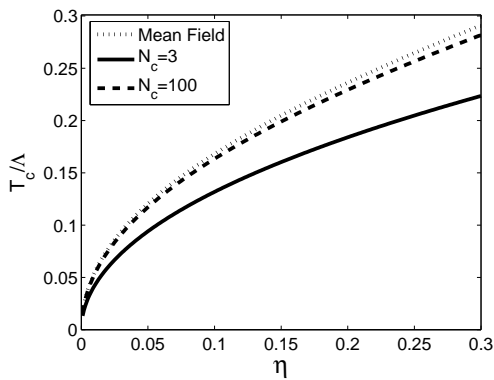}
\caption{The critical temperature $T_c$ of chiral phase transition
as a function of the coupling strength $\eta$ in mean field
approximation (dotted line) and in the case including pseudogap
effect with $N_c=3$ (solid line) and $100$ (dashed line). The
chemical potential $\mu$ is taken to be zero. \label{fig1}}
\end{center}
\end{figure}

The NJL model is non-renormalizable, the simplest regularization
scheme is to use a three momentum cutoff $\Lambda$ to regularize
the integrals over quark and pair momenta. In the vacuum, the
spontaneous breaking of chiral symmetry occurs when the coupling
exceeds the critical value $g_c=4\pi^2/(N_cN_f\Lambda^2)$. In the
following we use a dimensionless quantity $\eta=g/g_c-1$ to denote
the coupling strength. To fit the pion decay constant $f_\pi=94$
MeV and constituent quark mass $m=312$ MeV in the vacuum, we take
$\Lambda=653$ MeV and $g/4=5.01$ (GeV)$^{-2}$~\cite{zhuang} which
result in the physical coupling $\eta = 0.3$. In Fig.\ref{fig1} we
calculate the transition temperature $T_c$ as a function of the
coupling in mean field approximation and in the case including
pseudogap effect with $N_c=3$ and $100$. It is clear that the
pseudogap effect can be neglected at large enough $N_c$ and
disappears in the limit $N_c\rightarrow\infty$. For a finite value
of $N_c$, the transition temperature with contribution from pair
fluctuations is lower than the one in mean field approximation,
and the difference between the two increases with increasing
coupling. This conclusion is consistent with the results obtained
in other approaches, like the non-linear sigma model approach of
the NJL model~\cite{qqbar,qqbar2}.
\begin{figure}[!htb]
\begin{center}
\includegraphics[width=7cm]{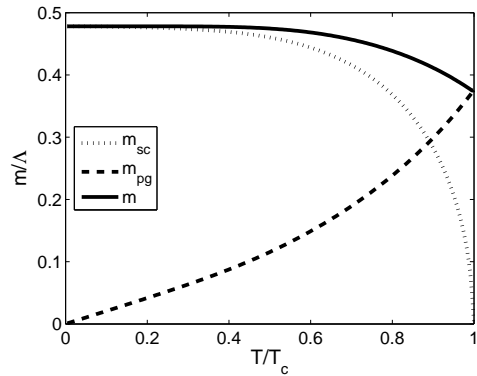}
\caption{The temperature dependence of the order parameter
$m_{\text{sc}}$ (dotted line), the pseudogap $m_{\text{pg}}$
(dashed line) and the total quark mass $m$ (solid line) at
$\mu=0$, $\eta=0.3$ and $N_c=3$. \label{fig2}}
\end{center}
\end{figure}

In the following we focus on the physical case of $N_c=3$ and
$\eta=0.3$. In Fig.\ref{fig2} we show the temperature dependence
of the order parameter $m_{\text{sc}}$, the pseudogap
$m_{\text{pg}}$ and the total quark mass $m$ at zero chemical
potential. At zero temperature, the pseudogap is zero and the
quark mass comes purely from the dynamical symmetry breaking. In
the low temperature domain $T<0.4\ T_c$, there is a good linear
relation between $m_{\text{pg}}$ and $T$. With increasing
temperature, while the order parameter is reduced, the pseudogap
goes up and dominates the quark mass near the transition
temperature. Due to the large pseudogap around the transition, the
effective quark mass at $T_c=146$ MeV is about $80\%$ of its value
in the vacuum. This is in contrary to the mean field picture where
the quark mass is equivalently considered as the order parameter
of chiral phase transition and then keeps zero in the chiral
symmetry restoration phase.
\begin{figure}[!htb]
\begin{center}
\includegraphics[width=7cm]{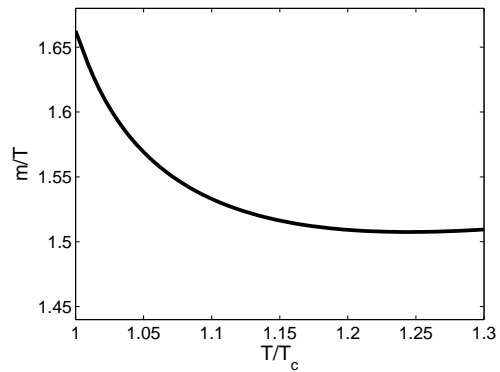}
\caption{The scaled quark mass $m/T$ as a function of scaled
temperature $T/T_c$ at $\mu=0$, $\eta=0.3$ and $N_c=3$ in the
region above but close to the critical temperature $T_c$.
\label{fig3}}
\end{center}
\end{figure}
\begin{figure}[!htb]
\begin{center}
\includegraphics[width=7cm]{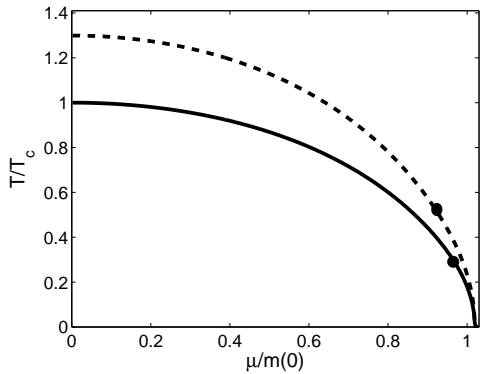}
\caption{The phase diagram in the $T-\mu$ plane with $\eta=0.3$
and $N_c=3$. The solid and dashed lines are, respectively, the
phase transition lines with and without considering pair
fluctuations. The black dots indicate the corresponding
tricritical points. \label{fig4}}
\end{center}
\end{figure}

Above the transition temperature $T_c$, the order parameter
$m_{\text{sc}}$ vanishes, but the pseudogap $m_{\text{pg}}$ is
generally not zero in the temperature region close to $T_c$. As we
mentioned above, the approximation (\ref{app}) is in principle not
valid above the transition temperature. However, for temperatures
not much larger than $T_c$, it can be used to predict the
quantitative behavior of pair fluctuations. In Fig.\ref{fig3}, we
calculate the quark mass $m$ in the temperature domain
$T_c<T<T_c^{\text{mf}}=1.3T_c$, where $T_c^{\text{mf}}=190$ MeV is
the critical temperature at $\mu=0$ in mean field approximation.
In this temperature region, the quark mass comes purely from the
pseudogap. For temperatures $T>1.2T_c$, there is approximately a
linear relation between the quark mass and temperature, $m\propto
T$. This pseudogap induced quark mass can explain the fact why any
thermodynamic function can not reach its Stefan-Boltzmann limit
found by lattice calculations~\cite{SB}.

We now turn to the case of finite chemical potential. The quark
chemical potential leads to a mismatch between the quark and
antiquark Fermi surfaces, like the magnetic field which results in
a Zeeman splitting between the spin-up and spin-down electrons in
a superconductor. The phase diagram in the $T-\mu$ plane is shown
in Fig.\ref{fig4} where $T_c=146$ MeV is the critical temperature
at $\mu=0$ and $m(0)=312$ MeV is the quark mass in the vacuum. The
critical temperature is reduced by pair fluctuations, but the
pseudogap effect on the phase transition decreases with increasing
chemical potential. At zero temperature, the two phase transition
lines with and without considering pair fluctuations coincide. In
both cases, the phase transition is of second order at low
chemical potential and first order at high chemical potentia. The
tricritical point is shifted from $(T,\ \mu)=(78,\ 285)$ MeV in
mean field approximation to $(45,\ 300)$ MeV when pair
fluctuations are taken into account. In between the two phase
transition lines, there exist quarks and pairs, like a
superconductor in a magnetic field~\cite{magnetic}.

It is necessary to note that, the pairs considered in the $G_0 G$
scheme are not the real collective excitation
modes~\cite{BCSBEC3}, they just reflect the beyond-BCS effect on
the fermion propagator. The real collective modes, especially the
Goldstone modes, are constructed by the dressed fermion
propagators $G_{\text{L,R}}$~\cite{BCSBEC3}.

In summary, based on the analogy between the mechanisms of chiral
symmetry breaking and superconducting, we have extended the
BCS-BEC crossover theory in the NJL model to including $\bar q q$
pair fluctuations at finite temperature and density. By using the
BEC condition in the symmetry breaking phase, the fluctuations are
treated as a quark pseudogap which reduces the critical
temperature of chiral phase transition. While the NJL model lacks
confinement and gluon degrees of freedom, the $\bar q q$ pair
fluctuations induce a large pseudogap above the critical
temperature which can be used to explain the large thermal quark
mass observed in lattice QCD.

{\bf Acknowledgement:} This work is supported by Grants
NSFC10575058, 10735040 and the 973-program 2007CB815000.

\end{document}